\documentclass[preprint,showpacs,preprintnumbers,amsmath,amssymb]{revtex4}
\usepackage{epsfig}
\usepackage{graphicx}% Include figure files
\usepackage{dcolumn}% Align table columns on decimal point
\usepackage{bm}% bold math
\usepackage{threeparttable}

\def\beq{\begin{equation}}
\def\eeq{\end{equation}}
\def\eeqn{\end{equation}}
\newcommand\iden{\leavevmode\hbox{\small1\normalsize\kern-.33em1}}

\newcommand{\bea} {\begin{eqnarray}}
\newcommand{\eea} {\end{eqnarray}}

\newcommand{\nn}{\nonumber}
\let\jnfont=\rm
\def\NPB#1,{{\jnfont Nucl.\ Phys.\ B }{\bf #1},}
\def\PLB#1,{{\jnfont Phys.\ Lett.\ B }{\bf #1},}
\def\EPJC#1,{{\jnfont Eur.\ Phys.\ Jour.\ C }{\bf #1},}
\def\PRD#1,{{\jnfont Phys.\ Rev.\ D }{\bf #1},}
\def\PRL#1,{{\jnfont Phys.\ Rev.\ Lett.\ }{\bf #1},}
\def\MPLA#1,{{\jnfont Mod.\ Phys.\ Lett.\ A }{\bf #1},}
\def\JPG#1,{{\jnfont J.\ Phys.\ G }{\bf #1},}
\def\CTP#1,{{\jnfont Commun.\ Theor.\ Phys.\ }{\bf #1},}
\def\JHEP#1,{{\jnfont JHEP \ }{\bf #1},}
\def\NPPS#1,{{\jnfont Nucl.\ Phys.\ Proc.\ Suppl.\ }{\bf #1},}
\def\CPC#1,{{\jnfont Comput.\ Phys.\ Commun.\ }{\bf #1},}
\def\CPL#1,{{\jnfont Chin.\ Phys.\ Lett. }{\bf #1},}
\def\APPB#1,{{\jnfont Acta\ Phys.\ Polon.\ B }{\bf #1},}

\def\lsim{\raise0.3ex\hbox{$<$\kern-0.75em\raise-1.1ex\hbox{$\sim$}}}
\def\gsim{\raise0.3ex\hbox{$>$\kern-0.75em\raise-1.1ex\hbox{$\sim$}}}

\begin{document}

\title{\ \\[10mm] Study of the heavy CP-even Higgs with mass 125 GeV
in two-Higgs-doublet models at the LHC and ILC}

\author{Lei Wang, Xiao-Fang Han}

\affiliation{ Department of Physics, Yantai University, Yantai
264005, PR China \vspace{0.5cm} }

%---------------------------------------------------------------------------

\begin{abstract}
We assume that the 125 GeV Higgs discovered at the LHC is the heavy
CP-even Higgs of the two-Higgs-doublet models, and examine the
parameter space in the Type-I, Type-II, Lepton-specific and Flipped
models allowed by the latest Higgs signal data, the relevant
experimental and theoretical constraints. Further, we show the
projected limits on $\tan\beta$, $\sin(\beta-\alpha)$, $Hf\bar{f}$
and $HVV$ couplings from the future measurements of the 125 GeV
Higgs at the LHC and ILC, including the LHC with integrated
luminosity of 300 fb$^{-1}$ (LHC-300 fb$^{-1}$) and 3000 fb$^{-1}$
(LHC-3000 fb$^{-1}$) as well as the ILC at $\sqrt{s}=250$ GeV
(ILC-250 GeV), $\sqrt{s}=500$ GeV (ILC-500 GeV) and $\sqrt{s}=1000$
GeV (ILC-1000 GeV). Assuming that the future Higgs signal data have
no deviation from the SM expectation, the LHC-300 fb$^{-1}$,
LHC-3000 fb$^{-1}$ and ILC-1000 GeV can exclude the wrong-sign
Yukawa coupling regions of the Type-II, Flipped and Lepton-specific
models at the $2\sigma$ level, respectively. The future experiments
at the LHC and ILC will constrain the Higgs couplings to be very
close to SM values, especially for the $HVV$ coupling.

\end{abstract}

\maketitle

\section{Introduction}
A 125 GeV Higgs boson has been discovered in the ATLAS and CMS
experiments at the LHC \cite{cmsh,atlh}. A number of new
measurements or updates of existing ones were presented in  ICHEP
2014 \cite{k-5,k-6}. Especially the diphoton signal strength is
changed from $1.6\pm0.4$ to $1.17\pm0.27$ for ATLAS \cite{k-7} and
from $0.78^{+0.28}_{-0.16}$ to $1.12^{+0.37}_{-0.32}$ for CMS
\cite{k-8}. There are some updates in the $ZZ$ \cite{k-9,k-10}, $WW$
\cite{k-11,k-12}, $b\bar{b}$ \cite{k-13}, $\tau\bar{\tau}$
\cite{k-14} decay modes, and the $t\bar{t}H$ events \cite{k-15,k-16}
from ATLAS and CMS, as well as an overall update from the D0
\cite{k-17} since 2013. The properties of this particle with large
experimental uncertainties agree with the Standard Model (SM)
predictions. The two-Higgs-doublet model (2HDM) has very rich Higgs
phenomenology, including two neutral CP-even Higgs bosons $h$ and
$H$, one neutral pseudoscalar $A$, and two charged Higgs $H^{\pm}$.
There are four traditional types for 2HDMs, Type-I \cite{i-1,i-2},
Type-II \cite{i-1,ii-2}, Lepton-specific, and Flipped models
\cite{xy-1,xy-2,xy-3,xy-4,xy-5,xy-6} according to their different
Yukawa couplings, in which the tree-level flavor changing neutral
currents (FCNC) are forbidden by a discrete symmetry. In addition,
there is no tree-level FCNC in the 2HDM that allows both doublets to
couple to the fermions with aligned Yukawa matrices \cite{a2hm-1}.
The recent Higgs data have been used to constrain these 2HDMs over
the last few months
\cite{2h-0,2h-1,2h-2,2h-3,2h-4,2h-5,2h-6,2h-7,2h-8,2h-9,2h-10,2h-11,2h-21,2h-12,2h-13,2h-14,
2h-15,2h-afb,2h-16,2h-17,2h-18,2h-19,2h-20,a2hw-1,a2hw-2,a2hw-3,a2hw-4,a2hw-5,a2hw-6,a2hw-7}.

In this paper, we assume that the 125 GeV Higgs discovered at the
LHC is respectively the heavy CP-even Higgs of the Type-I, Type-II,
Lepton-specific and Flipped 2HDMs, and examine the parameter space
allowed by the latest Higgs signal data, the non-observation of
additional Higgs at the collider, and the theoretical constraints
from vacuum stability, unitarity and perturbativity as well as the
experimental constraints from the electroweak precision data and
flavor observables. Further, we analyze how well 2HDMs can be
distinguished from SM by the future measurements of the 125 GeV
Higgs at the LHC and ILC, including the LHC with the center of mass
energy $\sqrt{s}=14$ TeV and integrated luminosity of 300 fb$^{-1}$
(LHC-300 fb$^{-1}$) and 3000 fb$^{-1}$ (LHC-3000 fb$^{-1}$) as well
as the ILC at $\sqrt{s}=250$ GeV (ILC-250 GeV), $\sqrt{s}=500$ GeV
(ILC-500 GeV) and $\sqrt{s}=1000$ GeV (ILC-1000 GeV). For the 125
GeV Higgs is the light CP-even Higgs, the projected limits on 2HDMs
from the future measurements of the 125 GeV Higgs at the LHC and ILC
have been studied in \cite{2h-13,2h-14}.

Our work is organized as follows. In Sec. II we recapitulate the
two-Higgs-doublet models. In Sec. III we introduce the numerical
calculations. In Sec. IV, we examine the implications of the latest
Higgs signal data on the 2HDMs and projected limits on the 2HDMs
from the future measurements of the 125 GeV Higgs at the LHC and ILC
after imposing the theoretical and experimental constraints.
Finally, we give our conclusion in Sec. V.

\section{two-Higgs-doublet models}
The Higgs potential with a softly broken $Z_2$ symmetry is written
as \cite{2h-poten}
\begin{eqnarray} \label{V2HDM} \mathrm{V} &=& m_{11}^2
(\Phi_1^{\dagger} \Phi_1) + m_{22}^2 (\Phi_2^{\dagger}
\Phi_2) - \left[m_{12}^2 (\Phi_1^{\dagger} \Phi_2 + \rm h.c.)\right]\nonumber \\
&&+ \frac{\lambda_1}{2}  (\Phi_1^{\dagger} \Phi_1)^2 +
\frac{\lambda_2}{2} (\Phi_2^{\dagger} \Phi_2)^2 + \lambda_3
(\Phi_1^{\dagger} \Phi_1)(\Phi_2^{\dagger} \Phi_2) + \lambda_4
(\Phi_1^{\dagger}
\Phi_2)(\Phi_2^{\dagger} \Phi_1) \nonumber \\
&&+ \left[\frac{\lambda_5}{2} (\Phi_1^{\dagger} \Phi_2)^2 + \rm
h.c.\right].
\end{eqnarray}
We focus on the CP-conserving model in which all $\lambda_i$ and
$m_{12}^2$ are real. The two complex scalar doublets have the
hypercharge $Y = 1$,
\begin{equation}
\Phi_1=\left(\begin{array}{c} \phi_1^+ \\
\frac{1}{\sqrt{2}}\,(v_1+\phi_1^0+ia_1)
\end{array}\right)\,, \ \ \
\Phi_2=\left(\begin{array}{c} \phi_2^+ \\
\frac{1}{\sqrt{2}}\,(v_2+\phi_2^0+ia_2)
\end{array}\right).
\end{equation}
Where the electroweak vacuum expectation values (VEVs) $v^2 = v^2_1
+ v^2_2 = (246~\rm GeV)^2$, and the ratio of the two VEVs is defined
as usual to be $\tan\beta=v_2 /v_1$. After spontaneous electroweak
symmetry breaking, there are  five mass eigenstates: two neutral
CP-even $h$ and $H$, one neutral pseudoscalar $A$, and two charged
scalar $H^{\pm}$.

The tree-level couplings of the neutral Higgs bosons can have
sizable deviations from those of SM Higgs boson. Table \ref{dlcoup}
shows the couplings of the heavy CP-even Higgs with respect to those
of the SM Higgs boson in the Type-I, Type-II, Lepton-specific and
Flipped models.

%%%%%%%%%%%%%%%%%%%%
\begin{table}
\caption{The tree-level couplings of the heavy CP-even Higgs with
respect to those of the SM Higgs boson. $u$, $d$ and $l$ denote the
up-type quarks, down-type quarks and the charged leptons,
respectively. } \vspace{0.5cm}
  \setlength{\tabcolsep}{2pt}
  \centering
  \begin{tabular}{|c|c|c|c|c|}
    \hline
     ~model&~$HVV$~$(WW,~ZZ)$~~& ~~~~$Hu\bar{u}$~~~~ &~~~~ $Hd\bar{d}$~~~~&~~~~ $Hl\bar{l}$~~~~\\
    \hline
     ~Type-I~
     & $\cos(\beta-\alpha)$ & $\frac{\sin\alpha}{\sin\beta}$
     & $\frac{\sin\alpha}{\sin\beta}$
     & $\frac{\sin\alpha}{\sin\beta}$
     \\
     ~Type-II~
      & $\cos(\beta-\alpha)$ &$\frac{\sin\alpha}{\sin\beta}$
     &$\frac{\cos\alpha}{\cos\beta}$
     &$\frac{\cos\alpha}{\cos\beta}$
     \\
     ~Lepton-specific~
     & $\cos(\beta-\alpha)$ & $\frac{\sin\alpha}{\sin\beta}$
     & $\frac{\sin\alpha}{\sin\beta}$
     & $\frac{\cos\alpha}{\cos\beta}$
     \\
     ~ Flipped~
     & $\cos(\beta-\alpha)$ & $\frac{\sin\alpha}{\sin\beta}$
     & $\frac{\cos\alpha}{\cos\beta}$
     & $\frac{\sin\alpha}{\sin\beta}$
     \\
     \hline
      \end{tabular}
\label{dlcoup}
\end{table}
%%%%%%%%%%%%%%%%%%%%%%%%%%%%%%%%%%%%%%%%%%%%%%%%%
\section{numerical calculations}
Using the method taken in
\cite{chi-1,chi-2,chi-3,susy-2,lh-1,lh-2,chi-4,chi-5}, we perform a
global fit to the latest Higgs data of 29 channels (see Tables I-V
in \cite{kmdata}). The signal strength for the $i$ channel is
defined as \beq\mu_i=\epsilon_{ggh}^i R_{ggH}+\epsilon_{VBF}^i
R_{VBF}+\epsilon_{VH}^i R_{VH}+\epsilon_{t\bar{t}H}^i
R_{t\bar{t}H}.\eeq Where $R_{j}=\frac{(\sigma \times
BR)_j}{(\sigma\times BR)_j^{SM}}$ with $j$ denoting the partonic
processes $ggH,~VBF,~VH,$ and $t\bar{t}H$. $\epsilon_{j}^i$ denotes
the assumed signal composition of the partonic process $j$, which
are given in Tables I-V of \cite{kmdata}. The $\chi^2$ for an
uncorrelated observable is \beq
\chi^2_i=\frac{(\mu_i-\mu^{exp}_i)^2}{\sigma_i^2},\eeq where
$\mu^{exp}_i$ and $\sigma_i$ denote the experimental central value
and uncertainty for the $i$ channel. The uncertainty asymmetry is
retained in our calculations. For the two correlated observables, we
use \beq \chi^2_{i,j}=\frac{1}{1-\rho^2}
\left[\frac{(\mu_i-\mu^{exp}_i)^2}{\sigma_i^2}+\frac{(\mu_j-\mu^{exp}_j)^2}{\sigma_j^2}
-2\rho\frac{(\mu_i-\mu^{exp}_i)}{\sigma_i}\frac{(\mu_j-\mu^{exp}_j)}{\sigma_j}\right],\eeq
where $\rho$ is the correlation coefficient. We sum over the
$\chi^2$ for the 29 channels, and pay particular attention to the
surviving samples with $\chi^2-\chi^2_{\rm min} \leq 6.18$, where
$\chi^2_{\rm min}$ denotes the minimum of $\chi^2$. These samples
correspond to the 95.4\% confidence level regions in any two
dimensional plane of the model parameters when explaining the Higgs
data (corresponding to be within $2\sigma$ range).
%****************************z****************
%%%%%%%%%%%%%%%%%%%%%%%%%%%%%%%%%%%%%%%%%%%%%%%%%
\begin{table}
\caption{Projected 1$\sigma$ sensitivities of channels for the LHC
operating $\sqrt{s}=14$ TeV. The 300~fb$^{-1}$ and 3000~fb$^{-1}$
sensitivities are taken from Ref. \cite{future-alt} for ATLAS and
Ref. \cite{future-cms} for CMS. The assumed signal composition is
taken from Ref. \cite{hs-3}.}
\begin{center}
\begin{threeparttable}[b]
 \begin{tabular}{lccrrrrr}
%\begin{tabular}{cccccccc}
\hline
Channel & \multicolumn{2}{c}{Projected 1$\sigma$ sensitivity} & \multicolumn{5}{c}{Assumed signal composition (\%)} \\
 &  300~fb$^{-1}$& 3000~fb$^{-1}$  & ~ggH~ & ~VBF~ &~ WH~ & ~ ZH~ &$t\bar{t}H$\\
\hline
ATL $(pp)\to h\to \gamma\gamma~\mbox{(0jet)}$~ & $   0.22$ & $0.20$ & $  91.6$ & $   2.7$ & $   3.2$ & $   1.8$ & $   0.6$\\ % 11 21 31 41 51
ATL $(pp)\to h\to \gamma\gamma~\mbox{(1jet)}$~ & $   0.37$ & $0.37$ &$  81.8$ & $  13.2$ & $   2.9$ & $   1.6$ & $   0.5$\\ % 11 21 31 41 51
ATL $(pp)\to h\to \gamma\gamma~\mbox{(VBF-like)}$~ & $   0.47$ & $0.21$&     $  39.2$ & $  58.4$ & $   1.4$ & $   0.8$ & $   0.3$\\ % 11 21 31 41 51
ATL $(pp)\to h\to \gamma\gamma~(VH\mbox{-like)}$~ & $   0.77$ & $0.26$& $   2.5$ & $   0.4$ & $  63.3$ & $  15.2$ & $  18.7$\\ % 11 21 31 41 51
ATL $(pp)\to h\to \gamma\gamma~(t\bar{t}H\mbox{-like})$~ & $   0.55$ & $0.21$ &$   0.0$ & $   0.0$ & $   0.0$ & $   0.0$ & $ 100.0$\\ % 51
ATL $(pp)\to h\to WW~\mbox{(0jet)}$~ & $   0.20$ &$0.19$& $  98.2$ & $   1.8$ & $   0.0$ & $   0.0$ & $   0.0$\\ % 12 22
ATL $(pp)\to h\to WW~\mbox{(1jet)}$~ & $   0.36$ &$0.33$& $  88.4$ & $  11.6$ & $   0.0$ & $   0.0$ & $   0.0$\\ % 12 22
ATL $(pp)\to h\to WW~\mbox{(VBF-like)}$~ & $   0.21$ & $0.12$& $   8.1$ & $  91.9$ & $   0.0$ & $   0.0$ & $   0.0$\\ % 12 22
ATL $(pp)\to h\to ZZ~\mbox{(ggF-like)}$~ & $   0.13$ & $0.12$ &$  88.7$ & $   7.2$ & $   2.0$ & $   1.4$ & $   0.7$\\ % 13 23 33 43 53
ATL $(pp)\to h\to ZZ~\mbox{(VBF-like)}$~ & $   0.34$ &$0.21$ & $  44.7$ & $  53.2$ & $   0.7$ & $   0.4$ & $   1.0$\\ % 13 23 33 43 53
ATL $(pp)\to h\to ZZ~(VH\mbox{-like)}$~ & $   0.32$ & $0.13$ &$  30.1$ & $   9.0$ & $  34.8$ & $  12.1$ & $  14.0$\\ % 13 23 33 43 53
ATL $(pp)\to h\to ZZ~(t\bar{t}H\mbox{-like)}$~ & $   0.46$ & $0.20$& $   8.7$ & $   1.7$ & $   1.7$ & $   3.1$ & $  84.8$\\ % 13 23 33 43 53
ATL $(pp)\to h\to Z\gamma$~ & $   1.47$ & $0.57$&$  87.6$ & $   7.1$ & $   3.1$ & $   1.7$ & $   0.6$\\ % 16 26 36 46 56
ATL $(pp)\to h\to \mu\mu$~ & $   0.47$ & $0.19$ &$  87.6$ & $   7.1$ & $   3.1$ & $   1.7$ & $   0.6$\\ % 18 28 38 48 58
ATL $(pp)\to h\to \mu\mu~(t\bar{t}H)$~ & $   0.73$ & $0.26$ & $   0.0$ & $   0.0$ & $   0.0$ & $   0.0$ & $ 100.0$\\ % 58
ATL $(pp)\to h\to \tau\tau~(\mbox{VBF-like})$~ & $   0.22$ & $0.19$ & $   19.8$ & $ 80.2$ & $   0.0$ & $   0.0$ & $   0.0$\\ % 24
\hline
CMS $(pp)\to h\to \gamma\gamma$~ & $   0.06$ & $0.04$ & $  87.6$ & $   7.1$ & $   3.1$ & $   1.7$ & $   0.6$\\ % 11 21 31 41 51
CMS $(pp)\to h\to WW$~ & $   0.06$ & $0.04$ & $  88.1$  &$   7.1$ & $   3.1$ & $   1.7$ & $   0.0$\\ % 12 22 32 42
CMS $(pp)\to h\to ZZ$~ & $   0.07$ & $0.04$ &$  88.1$ & $   7.1$ & $   3.1$ & $   1.7$ & $   0.0$\\ % 13 23 33 43
CMS $(pp)\to h\to Z\gamma$~ & $   0.62$ & $0.20$& $  87.6$ & $   7.1$ & $   3.1$ & $   1.7$ & $   0.6$\\ % 16 26 36 46 56
CMS $(pp)\to h\to bb$~ & $   0.11$ & $0.05$ & $   0.0$ & $   0.0$ & $  57.0$ & $  32.3$ & $  10.7$\\ % 35 45 55
CMS $(pp)\to h\to \mu\mu$~ & $   0.40$ & $0.20$ & $  87.6$ & $   7.1$ & $   3.1$ & $   1.7$ & $   0.6$\\ % 18 28 38 48 58
CMS $(pp)\to h\to \tau\tau$~ & $   0.08$ & $0.05$ &$  68.6$ & $  27.7$ & $   2.4$ & $   1.4$ & $   0.0$\\ % 14 24 34 44
\hline
 \end{tabular}
\end{threeparttable}
 \end{center}
  \label{lhc300}
\end{table}

We employ $\textsf{2HDMC-1.6.4}$ \cite{2hc-1} to implement the
theoretical constraints from the vacuum stability, unitarity and
coupling-constant perturbativity, and calculate the oblique
parameters ($S$, $T$, $U$) and $\delta\rho$, whose experimental data
are from Ref. \cite{stupara}. $\delta\rho$ has been precisely
measured to be very close to 1 via Z-pole precision observables,
which gives a strong constraint on the mass difference between
various Higgses in the 2HDMs. $\textsf{SuperIso-3.3}$ \cite{spriso}
is used to implement the constraints from flavor observables,
including $B\to X_s\gamma$ \cite{bsrdstv}, $B_s\to\mu^+\mu^-$
\cite{bsuu}, $B_u\to\tau\nu$ \cite{butv} and $D_s\to\tau\nu$
\cite{bsrdstv}. $\textsf{HiggsBounds-4.1.3}$ \cite{hb-1,hb-2} is
employed to implement the exclusion constraints from the neutral and
charged Higgses searches at LEP, Tevatron and LHC at 95\% confidence
level. The constrains from $\Delta m_{B_d}$ and $\Delta m_{B_s}$
\cite{deltamdms} are considered, which are calculated using the
formulas in \cite{deltmq}. In addition, $R_b$ is calculated by \beq
R_{b}\equiv(1+\frac{S_b^{SM}}{s_b^{SM}+\delta
s_b}C_b)^{-1}=R^{SM}_b(1+\frac{\delta
s_b}{s_b^{SM}})/(1+R^{SM}_b\frac{\delta s_b}{s_b^{SM}}), \eeq where
\beq
s_b^{SM}=[(\bar{g}_b^L-\bar{g}_b^R)^2+(\bar{g}_b^L+\bar{g}_b^R)^2]
(1+\frac{3\alpha}{4\pi}Q^2_b),~~~~\delta s_b=s_b-s_b^{SM}.\eeq We
take the SM value $R^{SM}_b=0.21550\pm0.00003$ \cite{rb-sm} and the
experimental data $R^{exp}_b=0.21629\pm0.00066$ \cite{rb-exp}.
Following the calculations of Ref. \cite{rb-cal1}, we can obtain the
contributions of the charged and neutral Higgses to the tree-level
couplings $\bar{g}_b^L$ and $\bar{g}_b^R$, and the QCD corrections
is included, whose expressions are given in Ref. \cite{rb-cal2}.

The measurement uncertainties of Higgs signal rates will be sizably
reduced at the LHC-300 fb$^{-1}$ and LHC-3000 fb$^{-1}$. The
projected 1$\sigma$ sensitivities for channels are shown in Table
\ref{lhc300}. The sensitivities of ATLAS include the current theory
systematic uncertainties, the statistical and experimental
systematic uncertainties. The sensitivities of ATLAS taken in Ref.
\cite{hs-3} does not include the theory uncertainty. Therefore, the
sensitivities of ATLAS in Table \ref{lhc300} differ considerably
from those in Ref. \cite{hs-3}. The sensitivities of CMS correspond
to Scenario 2, which extrapolates the analyses of 7 and 8 TeV data
to 14 TeV assuming the theory uncertainties will be reduced by a
factor of 2 while other uncertainties are reduced by a factor of
$1/\sqrt{{\cal L}}$. The assumed signal composition is taken from
Ref. \cite{hs-3}, which obtains the signal composition for ATLAS
from Refs. \cite{future-alt, future-alt2}, and assumes typical
values of the signal composition for CMS guided by present LHC
measurements since CMS does not provide the signal composition.

 Using the projected 1$\sigma$ sensitivities for
channels, we define \beq\chi^2=\sum_{i}\frac{(\epsilon_{ggh}^i
R_{ggH}+\epsilon_{VBF}^i R_{VBF}+\epsilon_{WH}^i
R_{WH}+\epsilon_{ZH}^i R_{ZH}+\epsilon_{t\bar{t}H}^i
R_{t\bar{t}H}-1)^2}{\sigma_i^2}.\eeq Where $R_{j}=\frac{(\sigma
\times BR)_j}{(\sigma\times BR)_j^{SM}}$ with $j$ denoting the
partonic processes $ggH,~VBF,~WH,~ZH$ and $t\bar{t}H$.
$\epsilon_{j}^i$ and $\sigma_i$ denote the assumed signal
composition of the partonic process $j$ and $1\sigma$ uncertainty
for the signal $i$, respectively. Thus, $\chi^2$ is used to
determine how well 2HDMs can be distinguished from the SM by the
future measurement of the 125 GeV Higgs at the LHC. In another
words, we assume the future Higgs signal data have no deviation from
the SM expectation, and estimate the limits on the 2HDMs using the
projected 1$\sigma$ uncertainties for channels at the LHC-300
fb$^{-1}$ and LHC-3000 fb$^{-1}$.

\begin{table}
\caption{Projected 1$\sigma$ sensitivities of channels for the ILC
operating at $\sqrt s=250$ GeV, 500 GeV and 1000 GeV with a
corresponding integrated luminosity of 250 fb$^{-1}$, 500 fb$^{-1}$
and 1000 fb$^{-1}$, respectively \cite{future-ilc}.}
\begin{center}
\begin{tabular}{|c|ccc|}\hline
Channel & ~250 GeV~ &~ 500 GeV ~& ~ 1 TeV~\\
\hline
$\mu_{Zh}$  & 2.6\% & 3.0\% & --\\
$\mu_{Zh}(b\bar b)$  &  1.2\% & 1.8\% & --\\
$\mu_{Zh}(c\bar c)$  & 8.3\% & 13\% & --\\
$\mu_{Zh}(gg)$ &  7.0\%  & 11\% & --\\
$\mu_{Zh}(WW)$ &6.4\%   & 9.2\% & --\\
$\mu_{Zh}(ZZ)$  & 18\% & 25\% & --\\
$\mu_{Zh}(\tau\tau)$  & 4.2\%  & 5.4\% & --\\
$\mu_{Zh}(\gamma\gamma)$  & 34\%  & 34\% & --\\
$\mu_{Zh}(\mu\mu)$  & 100\%  &  -- & --\\
$\mu_{WW}(b\bar b)$  &  10.5\% & 0.7\% & 0.5\%\\
$\mu_{WW}(c\bar c)$  &  -- & 6.2\% & 3.1\%\\
$\mu_{WW}(gg)$  &  -- & 4.1\% & 2.6\%\\
$\mu_{WW}(WW)$  &  -- & 2.4\% & 1.6\%\\
$\mu_{WW}(ZZ)$  &  -- & 8.2\% & 4.1\%\\
$\mu_{WW}(\tau\tau)$  &  -- & 9.0\% & 3.1\%\\
$\mu_{WW}(\gamma\gamma)$  &  -- & 23\% & 8.5\%\\
$\mu_{WW}(\mu\mu)$  &  -- & -- & 31\%\\
$\mu_{t\bar t}(b\bar b)$  &  -- & 28\% & 6.0\%\\
\hline
\end{tabular}
\end{center}
\label{ilc250}
\end{table}

On the other hand, the design center of mass energy at the
International Linear Collider (ILC) are 250 GeV and 500 GeV with a
possibility to upgrade to 1 TeV. For the Higgs measurements, the
beam polarizations are tuned to be $(e^-, e^+) = (-0.8,+0.3)$ at 250
GeV and 500 GeV as well as $(e^-, e^+) = (-0.8,+0.2)$ at 1 TeV. At
$\sqrt{s}=250$ GeV, an absolute measurement of the production cross
section can be performed from the $Z$ Higgsstrahlung near threshold.
The weak boson fusion process dominates over the $Z$ Higgsstrahlung
process at 500 GeV and 1000 GeV. The projected 1$\sigma$
sensitivities of channels at the ILC are shown in Table
\ref{ilc250}. Using the projected 1$\sigma$ sensitivities for
channels at the ILC, we define
\beq\chi^2=\sum_{i}\frac{(R_i-1)^2}{\sigma_i^2},\eeq where $R_i$ and
$\sigma_i$ represent the signal strength prediction from the 2HDMs
and the $1\sigma$ uncertainty for the signal $i$, respectively.

In our calculations, the input parameters are taken as $m^2_{12}$,
$\tan\beta$, $\sin(\beta-\alpha)$ and the physical Higgs masses
($m_h$, $m_H$, $m_A$, $m_{H^{\pm}}$). We fix $m_H$ as 125 GeV, and
scan randomly the parameters in the following ranges:
\begin{eqnarray}
&&20~{\rm GeV }\leq m_h\leq125~{\rm GeV},~~~~50 {\rm\  GeV} \leq m_A,~m_{H^\pm}  \leq 800  {\rm\  GeV},\nn\\
&&-0.7 \leq \sin(\beta-\alpha) \leq 0.7, \hspace{0.83cm} 0.1 \leq \tan \beta \leq 40,\nn\\
&&-(400~{\rm GeV})^2 \leq m^2_{12} \leq (400~{\rm GeV})^2.
\end{eqnarray}

\section{results and discussions}
In addition to that the theoretical constraints are satisfied, we
require the 2HDMs to explain the experimental data of flavor
observables and the electroweak precision data within 2$\sigma$
range, and fit the current Higgs signal data, the future LHC and ILC
data at the $2\sigma$ level.

\begin{figure}[tb]
%\begin{center}
 \epsfig{file=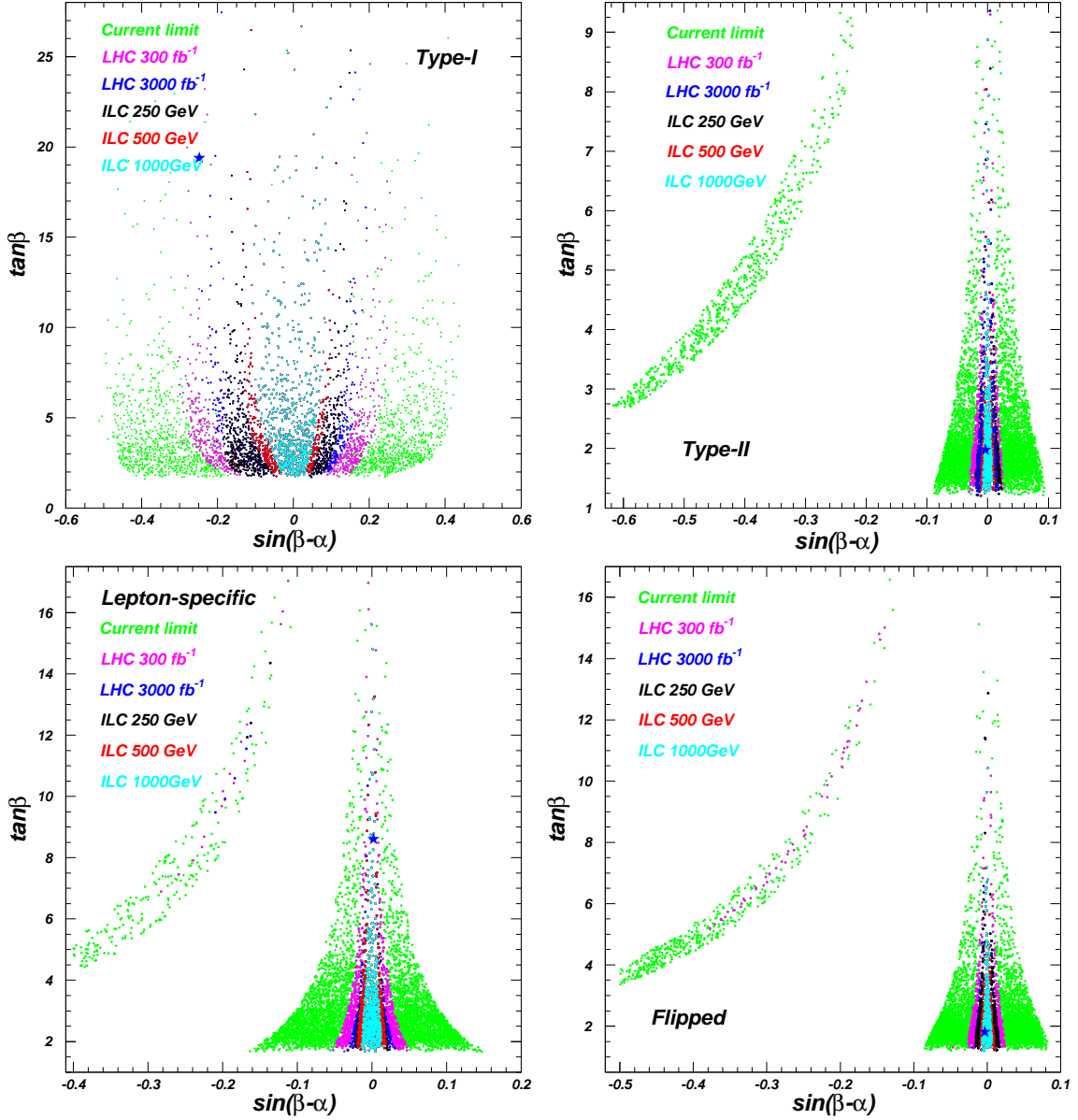,height=16.cm}
%\end{center}
\vspace{-0.3cm} \caption{The scatter plots of surviving samples
projected on the planes of $\sin(\beta-\alpha)$ versus $\tan\beta$.
The samples with the minimal values of $\chi^2$ are marked out as
stars.} \label{sba-tb}
\end{figure}

In Fig. \ref{sba-tb}, we project the surviving samples on the plane
of $\sin(\beta-\alpha)$ versus $\tan\beta$. $\tan\beta$ is required
to be larger than 1.6 for the Type-I and Lepton-specific models, and
1.1 for the Type-II and Flipped models. The main constraints are
from $\Delta m_{B_d}$ and $\Delta m_{B_s}$ which are sensitive to
$\cot\beta$. The Type-I model is less constrained than the other
three models by the current data. $\sin(\beta-\alpha)$ is allowed to
vary in the range of -0.55 and 0.5. In the Type-I model, the neutral
CP-even Higgs couplings to fermions have a universal varying factor.
In addition, the charged Higgs Yukawa couplings approach to zero in
the large $\tan\beta$ limit, which is less constrained by $B\to
X_s\gamma$ and $R_b$.

%%%%%%%%%%%%%%%%%%%%%
%%%%%%%%%%%%%%%%%%%%

Fig. \ref{sba-tb} shows that the surviving samples lie in the two
different regions in the Type-II, Lepton-specific and Flipped
models. In one region, the 125 GeV Higgs couplings are near the SM
values, called SM-like region. In the other region, at least one of
the Higgs Yukawa couplings has opposite sign to the corresponding
coupling to VV, called wrong-sign Yukawa coupling region. Now we
analyze the two regions in detail. In the four models, there are two
factors of $\frac{\cos\alpha}{\cos\beta}$ and
$\frac{\sin\alpha}{\sin\beta}$ for the heavy CP-even Higgs Yukawa
couplings normalized to the corresponding SM values.

For $\frac{\sin\alpha}{\sin\beta}$, \beq
\frac{\sin\alpha}{\sin\beta}=\cos(\beta-\alpha)-\sin(\beta-\alpha)\cot\beta.
\label{smss1}\eeq In the wrong-sign Yukawa coupling region where
both $\mid\varepsilon\mid$ and $\sin^2(\beta-\alpha)$ are much
smaller than 1,
 \beq \frac{\sin\alpha}{\sin\beta}=-1+\varepsilon,~~~\cos(\beta-\alpha)\simeq
1-\frac{1}{2}\sin(\beta-\alpha)^2. \label{smss2}\eeq From Eqs.
(\ref{smss1}) and (\ref{smss2}), we obtain \beq
\tan\beta=\frac{2\sin(\beta-\alpha)}{4-2\varepsilon-\sin^2(\beta-\alpha)}
\label{smss4}.\eeq This implies the wrong-sign $hf\bar{f}$ coupling
with a normalized factor $\frac{\sin\alpha}{\sin\beta}$ can only be
achieved for $\tan\beta$ is much smaller than 1, which is excluded
by the current experimental data as the above discussions.

For $\frac{\cos\alpha}{\cos\beta}$,
%%%%%%%%%%%%%%%%%%
 \beq
\frac{\cos\alpha}{\cos\beta}=\cos(\beta-\alpha)+\sin(\beta-\alpha)\tan\beta,
\label{smcc1}\eeq \beq
\frac{\cos\alpha}{\cos\beta}=\cos(\beta+\alpha)+\sin(\beta+\alpha)\tan\beta.
\label{wrongcc1}\eeq For $\cos(\beta-\alpha)=1$ and
$\cos(\beta+\alpha)=-1$, the $Hf\bar{f}$ couplings normalize to the
SM value equal to 1 and -1, which are the limiting cases of the
SM-like region and the wrong-sign Yukawa coupling region,
respectively.

In the wrong-sign Yukawa coupling region where both
$\mid\varepsilon\mid$ and $\sin^2(\beta-\alpha)$ are much smaller
than 1,
 \beq
\frac{\cos\alpha}{\cos\beta}=-1+\varepsilon,~~~\cos(\beta-\alpha)\simeq
1-\frac{1}{2}\sin(\beta-\alpha)^2. \label{wrongcc2}\eeq From Eqs.
(\ref{smcc1}) and (\ref{wrongcc2}), we obtain \beq
\tan\beta=\frac{\frac{1}{2}\sin(\beta-\alpha)^2+\varepsilon-2}{\sin(\beta-\alpha)}
\label{wrongcc3},\eeq
 \beq
\sin(\beta-\alpha)=\frac{\frac{1}{2}\sin(\beta-\alpha)^2+\varepsilon-2}{\tan\beta}
\label{wrongcc4}.\eeq From Eq. (\ref{wrongcc3}), the wrong-sign
$hf\bar{f}$ coupling with a normalized factor
$\frac{\cos\alpha}{\cos\beta}$ can only be achieved for $\tan\beta$
is much larger than 1 and $\sin(\beta-\alpha)<0$.

In the SM-like region, \beq
\frac{\cos\alpha}{\cos\beta}=1-\varepsilon,~~~\cos(\beta-\alpha)\simeq
1-\frac{1}{2}\sin(\beta-\alpha)^2. \label{smcc2}\eeq From Eqs.
(\ref{smcc1}) and (\ref{smcc2}), we obtain \beq \tan\beta
=\frac{\frac{1}{2}\sin(\beta-\alpha)^2-\varepsilon}{\sin(\beta-\alpha)}
\label{smcc3},\eeq
 \beq
\sin(\beta-\alpha)
=\frac{\frac{1}{2}\sin(\beta-\alpha)^2-\varepsilon}{\tan\beta}
\label{smcc4}.\eeq  Compared Eqs. (\ref{wrongcc3}) and
(\ref{smcc3}), the lower bound of $\tan\beta$ in the wrong-sign
Yukawa coupling region should be larger than that in the SM-like
region. Compared Eqs. (\ref{wrongcc4}) and (\ref{smcc4}), the
absolute value of $\sin(\beta-\alpha)$ in the wrong-sign Yukawa
coupling region should be larger than that in the SM-like region for
the same $\tan\beta$. Recently, Ref. \cite{2h-15} discusses the
wrong-sign Yukawa coupling of the light CP-even Higgs in the Type-II
model in detail.
%%%%%%%%%%%%%%%%%%%%%
\begin{figure}[tb]
%\begin{center}
 \epsfig{file=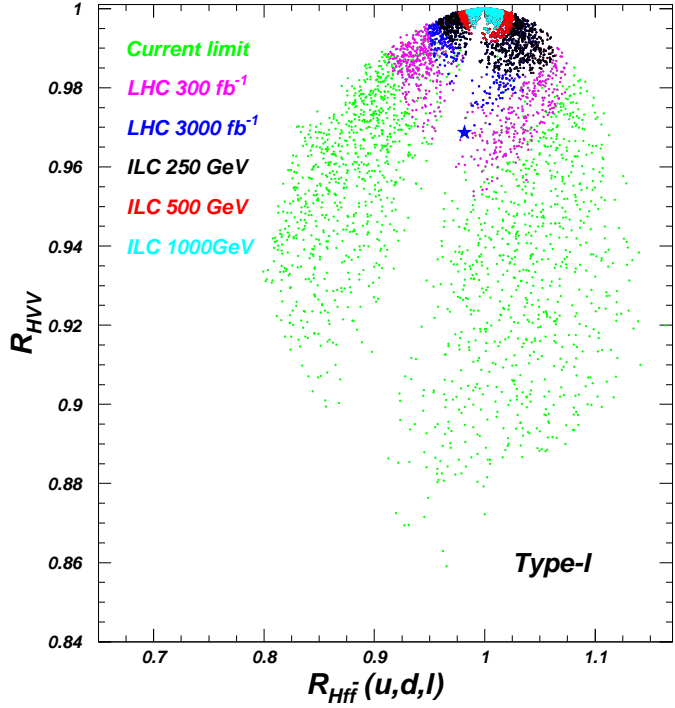,height=9.5cm}
%\end{center}
\vspace{-0.45cm} \caption{The scatter plots of surviving samples in
the Type-I model projected on the planes of $R_{Hf\bar{f}}~(u,d,l)$
versus $R_{HVV}$. Where $R_{Hf\bar{f}}$ and $R_{HVV}$ denote the
heavy CP-even Higgs couplings to $f\bar{f}$ and $VV$ normalized to
the corresponding SM values.} \label{coupi}
\end{figure}
%%%%%%%%%%%%%%%%%%%%

%%%%%%%%%%%%%%%%%%%%%
\begin{figure}[tb]
%\begin{center}
 \epsfig{file=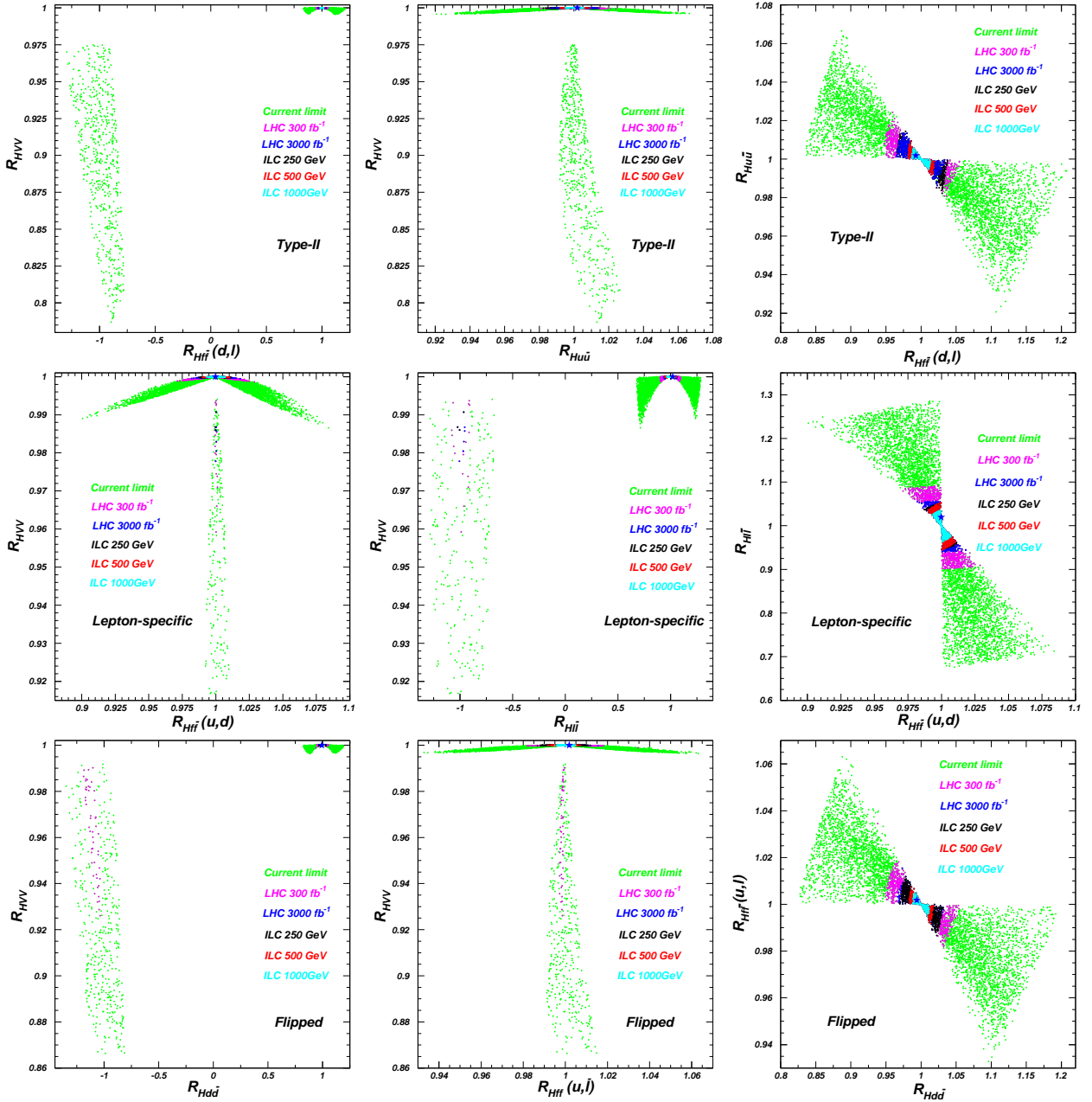,height=17.0cm}
 %\end{center}
\vspace{-1.1cm} \caption{Same as Fig. \ref{coupi}, but for the
Type-II, Lepton-specific and Flipped models.} \label{coupii}
\end{figure}
%%%%%%%%%%%%%%%%%%%%
Therefore, the wrong-sign Yukawa coupling can be achieved for the
$hd\bar{d}$ and $hl\bar{l}$ couplings in the Type-II model,
$hl\bar{l}$ in the Lepton-specific model, and $hd\bar{d}$ in the
Flipped model. The above analyses are confirmed by what are shown in
the Fig. \ref{sba-tb}. In the wrong-sign Yukawa coupling regions,
the current data require $\tan\beta>2.5$ for the Type-II model,
$\tan\beta>4$ for the Lepton-specific model and $\tan\beta>3$ for
the Flipped model. $\sin(\beta-\alpha)$ is allowed to be as low as
-0.62 for the Type-II model, -0.4 for the Lepton-specific model and
-0.5 for the Flipped model. In the SM-like regions, the current data
require $-0.18<\sin(\beta-\alpha)<0.16$ for the Lepton-specific
model, and $-0.1<\sin(\beta-\alpha)<0.1$ for the Type-II model and
the Flipped model.

For the Type-I model, the LHC-300 fb$^{-1}$, LHC-3000 fb$^{-1}$,
ILC-250 GeV, ILC-500 GeV and ILC-1000 GeV will gradually narrow the
allowed range of $\sin(\beta-\alpha)$. For the Type-II and Flipped
models, the LHC-300 fb$^{-1}$ can narrow the ranges of
$\sin(\beta-\alpha)$ sizably, and the ILC-250 GeV can not narrow the
ranges of $\sin(\beta-\alpha)$ more visibly than LHC-3000 fb$^{-1}$.

In Fig. \ref{coupi} and Fig. \ref{coupii}, we project the surviving
samples on the planes of the 125 GeV Higgs couplings. From Fig.
\ref{coupi}, for the Type-I model, we find that the allowed ranges
of $R_{HVV}$ and $R_{Hf\bar{f}}$ are $0.86\sim1.0$ and $0.8\sim1.17$
for the current constraints, $0.952\sim1.0$ and $0.911\sim1.075$ for
the LHC-300 fb$^{-1}$, $0.97\sim1.0$ and $0.948\sim1.048$ for the
LHC-3000 fb$^{-1}$, $0.983\sim1.0$ and $0.957\sim1.063$ for the
ILC-250 GeV, $0.991\sim1.0$ and $0.977\sim1.026$ for the ILC-500 GeV
as well as $0.994\sim1.0$ and $0.984\sim1.017$ for the ILC-1000 GeV.

For the Type-II model, in the wrong-sign $Hd\bar{d}$ and $Hl\bar{l}$
couplings region, the current data require $0.785<R_{HVV}<0.975$,
$-1.3<R_{Hd\bar{d}}~(R_{Hl\bar{l}})<-0.775$ and
$0.991<R_{Hu\bar{u}}<1.027$. The LHC-300 fb$^{-1}$ can exclude the
wrong-sign $Hd\bar{d}$ and $Hl\bar{l}$ couplings region at the
$2\sigma$ level. In the SM-like region, the current data require
$0.995<R_{HVV}<1.0$, $0.83<R_{Hd\bar{d}}~(R_{Hl\bar{l}})<1.22$ and
$0.92<R_{Hu\bar{u}}<1.07$. The future LHC and ILC experiments will
require $R_{HVV}$ to be very close to 1. The allowed ranges of
$R_{Hd\bar{d}}~(R_{Hl\bar{l}})$ and $R_{Hu\bar{u}}$ are
 $0.946\sim1.055$ and
$0.979\sim1.025$ for the LHC-300 fb$^{-1}$, $0.965\sim1.034$ and
$0.986\sim1.014$ for the LHC-3000 fb$^{-1}$, $0.965\sim1.038$ and
$0.981\sim1.015$ for the ILC-250 GeV, $0.981\sim1.019$ and
$0.99\sim1.009$ for the ILC-500 GeV as well as $0.986\sim1.014$ and
$0.993\sim1.006$ for the ILC-1000 GeV.

For the Lepton-specific model, in the wrong-sign $Hl\bar{l}$
coupling region, the current data require $0.915<R_{HVV}<0.995$,
$-1.3<R_{Hl\bar{l}}<-0.675$ and
$0.992<R_{Hu\bar{u}}~(R_{Hd\bar{d}})<1.01$. The LHC-300 fb$^{-1}$,
LHC-3000 fb$^{-1}$ and ILC-250 GeV can gradually constrain the
absolute values of Higgs couplings to $f\bar{f}$ and $VV$ to be
close to SM values in the wrong-sign $Hl\bar{l}$ coupling region,
and the ILC-1000 GeV can exclude the whole wrong-sign $Hl\bar{l}$
coupling region at the $2\sigma$ level. In the SM-like region, the
current data require $0.986<R_{HVV}<1.0$,
$0.675<R_{Hl\bar{l}}<1.288$ and
$0.9<R_{Hu\bar{u}}~(R_{Hd\bar{d}})<1.085$. The future LHC-300
fb$^{-1}$ will require $R_{HVV}$ to be in the range of 0.998 and
1.0. The other future LHC and ILC experiments will require $R_{HVV}$
to be very close to 1. The allowed ranges of
$R_{Hu\bar{u}}~(R_{Hd\bar{d}})$ and $R_{Hl\bar{l}}$ are
 $0.97\sim1.03$ and
$0.901\sim1.091$ for the LHC-300 fb$^{-1}$, $0.982\sim1.018$ and
$0.94\sim1.058$ for the LHC-3000 fb$^{-1}$, $0.988\sim1.013$ and
$0.946\sim1.051$ for the ILC-250 GeV, $0.991\sim1.01$ and
$0.945\sim1.053$ for the ILC-500 GeV as well as $0.993\sim1.007$ and
$0.963\sim1.037$ for the ILC-1000 GeV.

For the Flipped model, in the wrong-sign $Hd\bar{d}$ coupling
region, the current data require $0.865<R_{HVV}<0.993$,
$-1.35<R_{Hd\bar{d}}<-0.81$ and
$0.991<R_{Hu\bar{u}}~(R_{Hl\bar{l}})<1.015$. The LHC-300 fb$^{-1}$
can exclude some samples with $R_{Hd\bar{d}}<-1$ and $R_{Hu\bar{u}}
$ very close to 1. The LHC-3000 fb$^{-1}$ can exclude the whole
wrong-sign $Hd\bar{d}$ coupling region at the $2\sigma$ level. In
the SM-like region, the current data require $0.996<R_{HVV}<1.0$,
$0.825<R_{Hd\bar{d}}<1.195$ and
$0.932<R_{Hu\bar{u}}~(R_{Hl\bar{l}})<1.064$. The future LHC and ILC
experiments will require $R_{HVV}$ to be very close to 1. The
allowed ranges of $R_{Hd\bar{d}}$ and
$R_{Hu\bar{u}}~(R_{Hl\bar{l}})$ are
 $0.946\sim1.056$ and
$0.981\sim1.018$ for the LHC-300 fb$^{-1}$, $0.965\sim1.034$ and
$0.988\sim1.015$ for the LHC-3000 fb$^{-1}$, $0.97\sim1.032$ and
$0.986\sim1.013$ for the ILC-250 GeV, $0.983\sim1.018$ and
$0.992\sim1.008$ for the ILC-500 GeV as well as $0.987\sim1.013$ and
$0.994\sim1.005$ for the ILC-1000 GeV.

%%%%%%%%%%%%%%%%%%%%%
\begin{figure}[tb]
%\begin{center}
 \epsfig{file=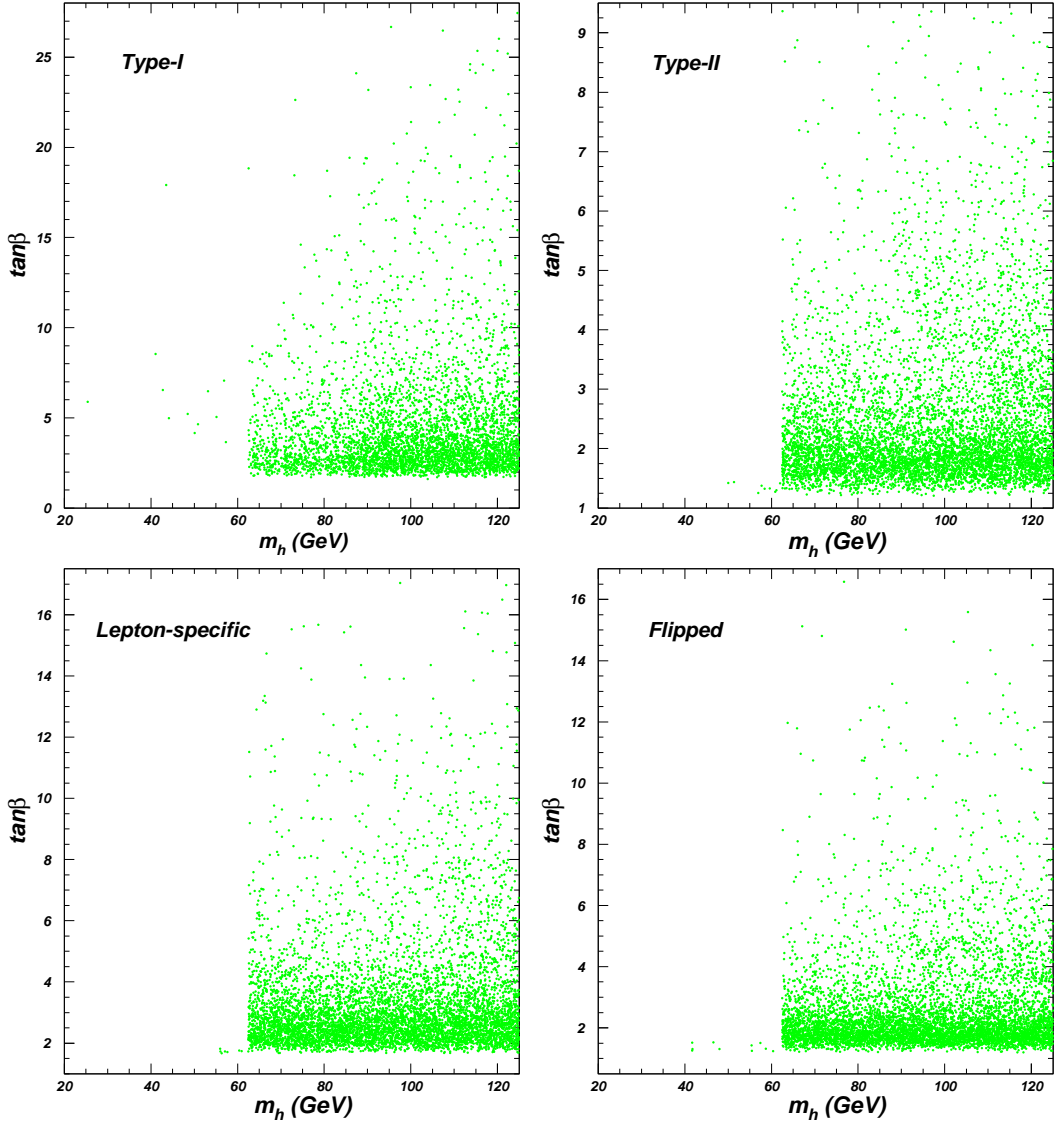,height=15.0cm}
%\end{center}
\vspace{-0.3cm} \caption{The scatter plots surviving the current
limits projected on the planes of $m_h$ versus $\tan\beta$.}
\label{mh-tb}
\end{figure}
%%%%%%%%%%%%%%%%%%%%

%%%%%%%%%%%%%%%%%%%%%
\begin{figure}[tb]
%\begin{center}
 \epsfig{file=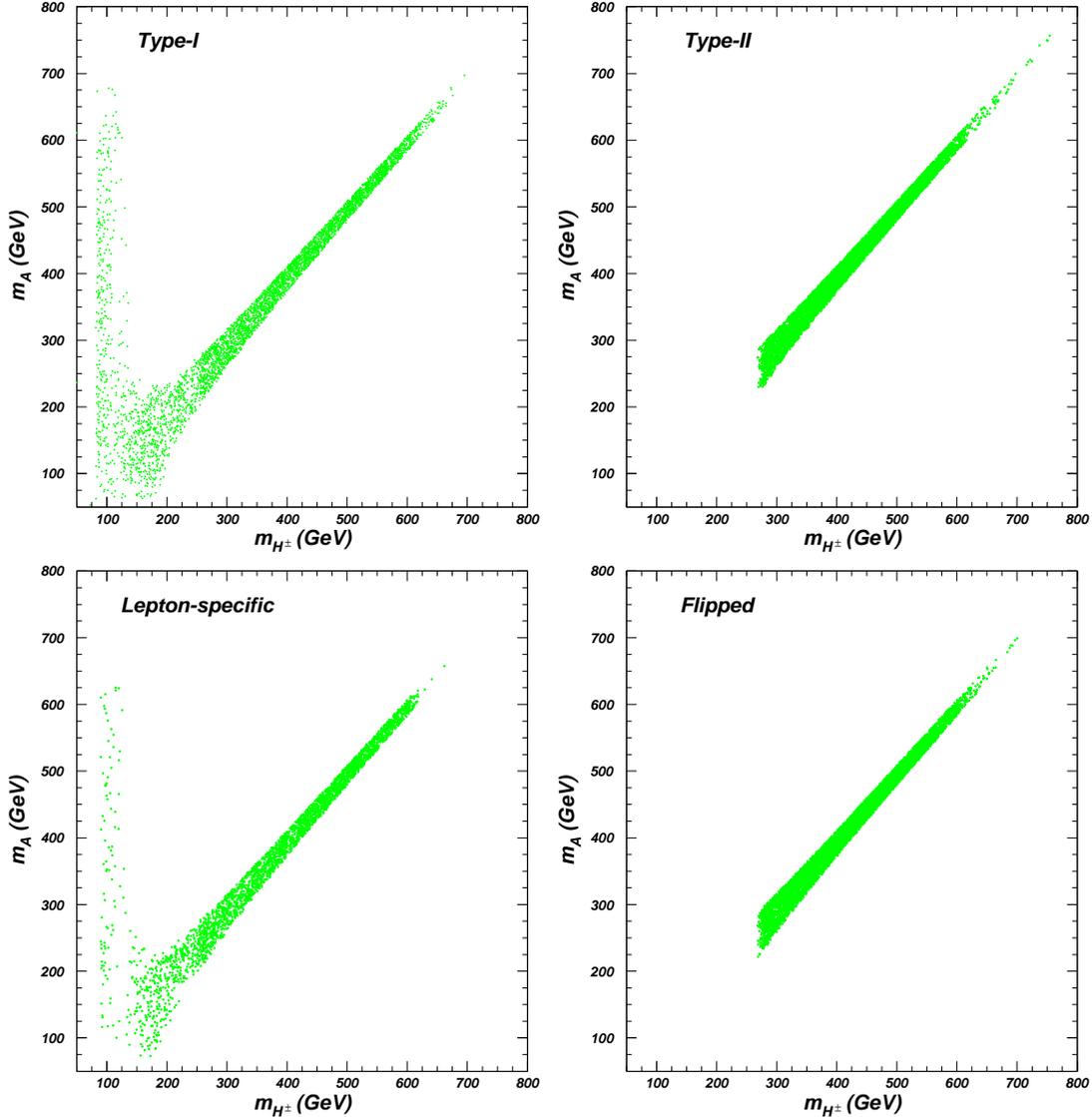,height=15.0cm}
%\end{center}
\vspace{-0.3cm} \caption{Same as Fig. \ref{mh-tb}, but projected on
the planes of $m_{H^{\pm}}$ versus $m_A$.} \label{mhp-ma}
\end{figure}
%%%%%%%%%%%%%%%%%%%%

%%%%%%%%%%%%%%%%%%%%%
\begin{figure}[tb]
%\begin{center}
 \epsfig{file=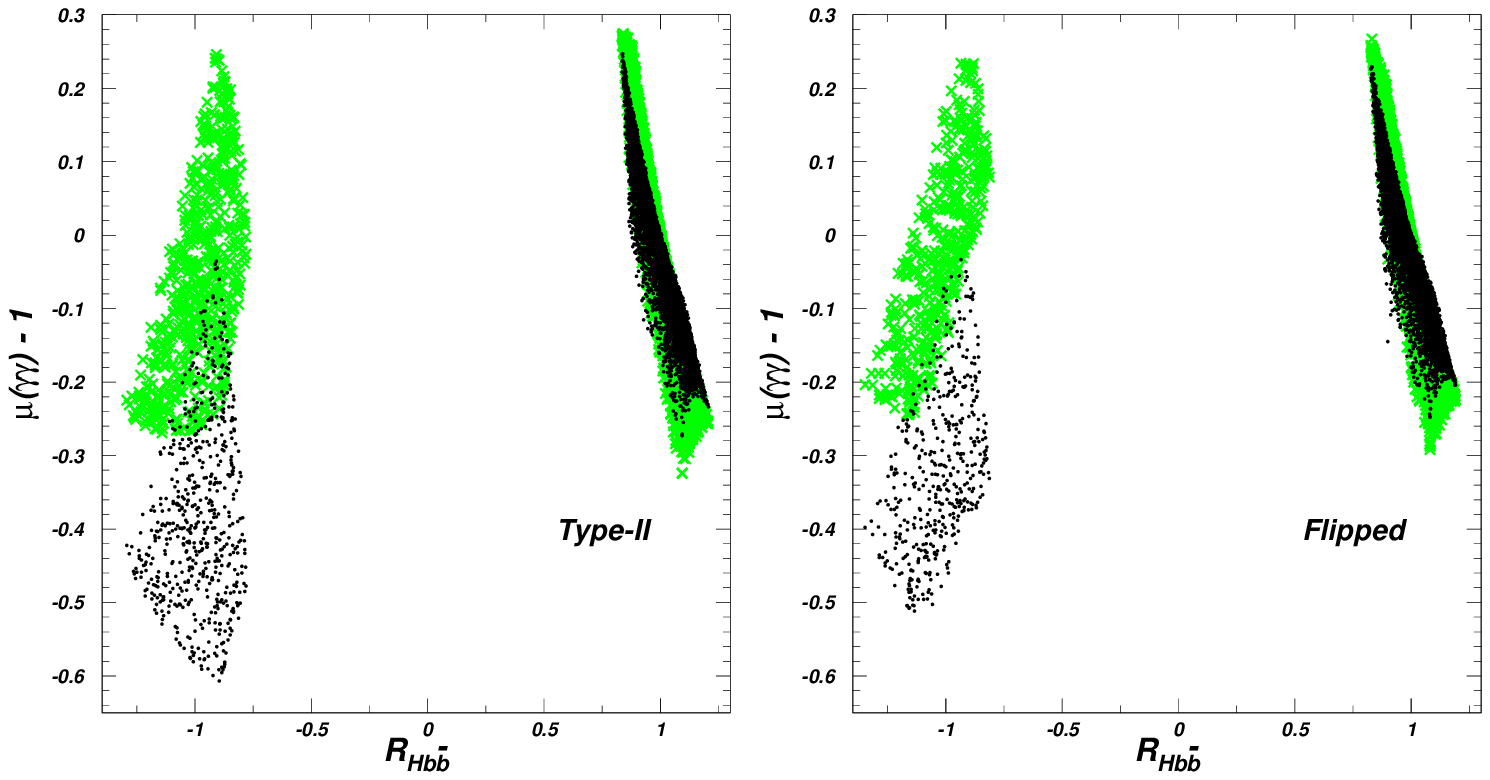,height=8.0cm}
%\end{center}
\vspace{-0.3cm} \caption{The scatter plots surviving the current
limits projected on the planes of $R_{Hb\bar{b}}$ versus the
diphoton Higgs signal at the LHC and ILC. The crosses (green) denote
the inclusive diphonon Higgs signal at the LHC, and the plots
(black) denote the diphoton Higgs signal via $Z$ Higgsstrahlung or
$WW$ fusion at the ILC.} \label{hrr}
\end{figure}
%%%%%%%%%%%%%%%%%%%%

Now we examine the allowed mass ranges of the light CP-even Higgs,
pseudoscalar and charged Higgs with the heavy CP-even Higgs being
the 125 GeV Higgs. Since the focus of this paper is studying the
limits on the heavy CP-even Higgs with mass 125 GeV at the current
and future collider, the projected limits on $m_h$, $m_A$ and
$m_{H^{\pm}}$ from the future collider are beyond the scope of this
paper. Therefore, we only show the mass ranges of $m_h$, $m_A$ and
$m_{H^{\pm}}$ allowed by the current limits in Fig. \ref{mh-tb} and
Fig. \ref{mhp-ma}. Since the decay $H\to hh$ is open for $m_h<$ 62.5
GeV, the $BR(H\to hh)$ has to be small enough that $H$ can fit the
LHC Higgs signal data at an adequate level. As a result, we only
obtain a few scattering of points for $m_h<$ 62.5 GeV in the Type-I,
Type-II, Lepton-specific, and Flipped models, respectively, as shown
in Fig. \ref{mh-tb}. If a very "fine-tuned" scan is employed, the
more low-$m_h$ points may be obtained.

Fig. \ref{mhp-ma} shows that $m_{H^{\pm}}$ is required to be larger
than 250 GeV in the Type-II and Flipped models due to the
constraints from the low energy flavor observables. There is small
mass difference between $m_A$ and $m_{H^{\pm}}$ mainly due to the
constraints of $\Delta\rho$. Since there is small mass difference
between $m_h$ and $m_H$, $m_A$ and $m_{H^{\pm}}$ should have the
small mass difference to cancel the contributions of $m_h$ and $m_H$
to $\Delta\rho$. In the Type-I and Lepton-specific models, since the
charged Higgs Yukawa couplings are suppressed by
$\frac{1}{\tan\beta}$, $m_{H^{\pm}}$ is allowed to be smaller than
100 GeV. Further, for $m_{H^{\pm}}$ is around $m_H$, the
contributions to $\Delta\rho$ from $(m_h,~m_{H^{\pm}})$ and
$(m_A,~m_{H^{\pm}})$ loops can be canceled by the $(m_h,~m_{H})$ and
$(m_A,~m_{H})$ loops. Thus $m_A$ is allowed to have large mass
difference from $m_{H^{\pm}}$ for $m_{H^{\pm}}$ is around 100 GeV.

Ref. \cite{1307.5973} shows that the second light Higgs boson
explanation of 125 GeV in the MSSM is ruled out by the present
experiments. Compared to Type-II model, the five Higgs masses in the
MSSM are not independent. Taking the mass of the second light Higgs
boson as 125 GeV, the mass of charged Higgs should be smaller than
200 GeV, which is excluded by the current experimental constraints,
especially for $BR(B\to X_{s}\gamma)$. Similarly, the current
experimental constraints require $m_{H^{\pm}}>$ 250 GeV in the
Type-II model. However, the Higgs masses in the Type-II model are
independent, and we can take enough large $m_{H^{\pm}}$ to avoid the
current experimental constraints.

For the wrong-sign Yukawa coupling of $b$-quark, the interference
between the $b$-quark and top-quark loops can give an enhanced
contribution to the effective coupling $hgg$, and the interference
between the $b$-quark and $W$ boson loops can give a suppressed
contribution to the effective coupling $h\gamma\gamma$. In Fig.
\ref{hrr}, we show the inclusive diphoton Higgs signal strength at
the LHC and the diphoton Higgs signal strength via $Z$
Higgsstrahlung and $WW$ fusion at the ILC (The diphoton Higgs signal
strength in the 2HDMs is the same for the $Z$ Higgsstrahlung and
$WW$ fusion processes at the ILC). The diphoton Higgs rate at the
ILC for $R_{Hb\bar{b}} < 0$ is sizably smaller than those for
$R_{Hb\bar{b}} > 0$. According to the projected sensitivities of
diphoton signal shown in the Table \ref{ilc250}, the diphoton Higgs
rates are within $2\sigma$ range of ILC-250 GeV -1.3
$<R_{Hb\bar{b}}<$ 1.2, and ILC-500 GeV for $R_{Hb\bar{b}}>$ 0, and
the ILC-1000 GeV can probe the wrong-sign Yukawa coupling of
$b$-quark in the Type-II and Flipped models by measuring the
diphoton Higgs signal via $WW$ fusion at $2\sigma$ level. By
measuring the inclusive diphoton Higgs signal at the LHC-300
fb$^{-1}$, CMS can detect the wrong-sign Yukawa coupling of Type-II
model and Flipped model at $2\sigma$ level.

Assuming the light CP-even Higgs is the discovered 125 GeV Higgs,
Ref. \cite{2h-13} shows $\tan\beta$ and $\cos(\beta-\alpha)$ within
$2\sigma$ ranges of the current Higgs data and the projected limits
from the future collider. Similar to the heavy CP-even Higgs, the
wrong-sign Yukawa coupling is absent in the Type-I model, and can
appear in the Type-II, Lepton-specific and Flipped models for
$\tan\beta>3$. For the Type-II, Lepton-specific and Flipped models,
$\cos(\beta-\alpha)$ is strongly constrained in the SM-like region,
and $\cos(\beta-\alpha)$ in the wrong-sign Yukawa coupling region is
allowed to be much larger than that in the SM-like region. The
current Higgs data allow $\cos(\beta-\alpha)$ to be as large as 0.55
for the Type-I, Type-II and Flipped models, and 0.5 for the
Lepton-specific model. The ILC-1000 GeV can give the strongest
constraints on $\cos(\beta-\alpha)$,
$\mid\cos(\beta-\alpha)\mid<0.4\%$ for the Type-II, Lepton-specific
and Flipped models as well as $\mid\cos(\beta-\alpha)\mid<8\%$ for
the Type-I model. For the heavy CP-even Higgs as the 125 GeV Higgs,
this paper shows that the ILC-1000 GeV gives the similar constraints
on $\sin(\beta-\alpha)$, $\mid \sin(\beta-\alpha)\mid <10\%$ for the
Type-I model, $\mid \sin(\beta-\alpha)\mid <0.8\%$ for the Type-II
model and Flipped models, and $\mid \sin(\beta-\alpha)\mid <1.4\%$
for the Lepton-specific model. This leads to that $R_{HVV}$ is very
close to 1 due to
$R_{HVV}=\cos(\beta-\alpha)\simeq1-\frac{1}{2}\sin(\beta-\alpha)^2$.

\section{Conclusion}
In this paper, we assume the 125 GeV Higgs discovered at the LHC is
the heavy CP-even Higgs of the Type-I, Type-II, Lepton-specific and
Flipped 2HDMs, and examine the parameter space allowed by the latest
Higgs signal data, the non-observation of additional Higgs at the
collider, and the theoretical constraints from vacuum stability,
unitarity and perturbativity as well as the experimental constraints
from the electroweak precision data and flavor observables. We
obtain the following observations:

(i) The current theoretical and experimental constraints favor a
small $\tan\beta$, but give a lower limit of $\tan\beta$,
$\tan\beta>1.6$ for the Type-I model, $\tan\beta>1.1$ (2.5) for the
SM-like region (wrong-sign Yukawa coupling region) of the Type-II
model, $\tan\beta>1.6$ (4.0) for the SM-like region (wrong-sign
Yukawa coupling region) of the Lepton-specific model, and
$\tan\beta>1.1$ (3.0) for the SM-like region (wrong-sign Yukawa
coupling region) of the Flipped model.

(ii) For the Type-I model, the current experimental data require
$0.86<R_{HVV}<1.0$ and $0.8<R_{Hf\bar{f}}~(u,d,l)<1.17$.

(iii) For the Type-II model, the current experimental data require
$0.785<R_{HVV}<0.975$, $-1.3<R_{Hd\bar{d}}~(R_{Hl\bar{l}})<-0.775$
and $0.991<R_{Hu\bar{u}}<1.027$ in the wrong-sign $Hd\bar{d}$ and
$Hl\bar{l}$ couplings region, and $0.995<R_{HVV}<1.0$,
$0.83<R_{Hd\bar{d}}~(R_{Hl\bar{l}})<1.22$ and
$0.92<R_{Hu\bar{u}}<1.07$ in the SM-like region.

(iv) For the Lepton-specific model, the current experimental data
require $0.915<R_{HVV}<0.995$, $-1.3<R_{Hl\bar{l}}<-0.675$ and
$0.992<R_{Hu\bar{u}}~(R_{Hd\bar{d}})<1.01$ in the wrong-sign
$Hl\bar{l}$ coupling region, and $0.986<R_{HVV}<1.0$,
$0.675<R_{Hl\bar{l}}<1.288$ and
$0.9<R_{Hu\bar{u}}~(R_{Hd\bar{d}})<1.085$ in the SM-like region.

(v) For the Flipped model, the current experimental data require
$0.865<R_{HVV}<0.993$, $-1.35<R_{Hd\bar{d}}<-0.81$ and
$0.991<R_{Hu\bar{u}}~(R_{Hl\bar{l}})<1.015$ in the wrong-sign
$Hd\bar{d}$ coupling region, and $0.996<R_{HVV}<1.0$,
$0.825<R_{Hd\bar{d}}<1.195$ and
$0.932<R_{Hu\bar{u}}~(R_{Hl\bar{l}})<1.064$ in the SM-like region.

Further, we give the projected limits on $\tan\beta$,
$\sin(\beta-\alpha)$, $Hf\bar{f}$ and $HVV$ couplings from the
future measurements of the 125 GeV Higgs at the LHC and ILC,
including the LHC-300 fb$^{-1}$, LHC-3000 fb$^{-1}$, ILC-250 GeV,
ILC-500 GeV and ILC-1000 GeV. Assuming that the future Higgs signal
data have no deviation from the SM expectation, the LHC-300
fb$^{-1}$, LHC-3000 fb$^{-1}$ and ILC-1000 GeV can exclude the
wrong-sign Yukawa coupling regions of the Type-II, Flipped and
Lepton-specific models at the $2\sigma$ level, respectively. The
future experiments at the LHC and ILC will constrain the Higgs
couplings to be very close to SM values, especially for the $HVV$
coupling.

\section*{Acknowledgment}
This work was supported by the National Natural Science Foundation
of China (NNSFC) under grant No. 11105116.

\end{document}